\documentclass[11pt]{article}
\usepackage[english]{babel}
\usepackage{tabu}
\usepackage[table]{xcolor}
\usepackage{amsmath}
\usepackage{amsfonts}
\usepackage{amssymb}
\usepackage[labelfont=bf]{caption}
\usepackage{paralist}
\usepackage[numbers,sort&compress]{natbib}
\usepackage{graphicx}

\usepackage{geometry}
\usepackage{times}

\usepackage{color}
\usepackage{wrapfig}
\usepackage{bibentry}
\usepackage{soul}

\begin{document}

\noindent\LARGE{\textbf{An introduction to linear stability analysis for deciphering spatial patterns in signaling networks}}
\vspace{0.6cm}

\noindent\large{\textbf{Jasmine A. Nirody \textit{$^{a}$} and Padmini Rangamani$^{\ast}$\textit{$^{b}$}}}\vspace{0.5cm}

\noindent \normalsize{Mathematical modeling is now used commonly in the analysis of signaling networks. With advances in high resolution microscopy, the spatial location of different signaling molecules and the spatio-temporal dynamics of signaling microdomains are now widely acknowledged as key features of biochemical signal transduction. Reaction-diffusion mechanisms are commonly used to model such features, often with a heavy reliance on numerical simulations to obtain results. However, simulations are parameter dependent and may not be able to provide an understanding of the full range of the system responses. Analytical approaches on the other hand provide a framework to study the entire phase space. In this tutorial, we provide a largely analytical method for studying reaction-diffusion models and analyzing their stability properties. Using two representative biological examples, we demonstrate how this approach can guide experimental design. In order to make such analyses more accessible, we provide an easy-to-use graphical user interface to test the stability behavior of biological systems (available at http://www.ocf.berkley.edu/$\sim$jnirody)$^\dag$.}
\vspace{0.5cm}

\clearpage
\section{Introduction}
\let\thefootnote\relax\footnote{\dag~Interactive Mathematica notebook for stability analysis of reaction diffusion equations at http://www.ocf.berkeley.edu/$\sim$jnirody.}
\footnotetext{\textit{$^{a}$~Biophysics Graduate Group, University of California, Berkeley, Berkeley, CA 94720}}
\footnotetext{\textit{$^{b}$~Department of Mechanical and Aerospace Engineering, University of California, San Diego, San Diego, CA 92093. Email: padmini.rangamani@eng.ucsd.edu}}

Mathematical models have proven to be extremely useful in elucidating the properties of biological signaling networks \citep{kholodenko1999quantification,markevich2004signaling,chen2010classic}. A common method of modeling such systems is to use \textit{ordinary differential equations} (ODEs) to describe the dynamics of various signaling components. However, given the compartmental nature of cells, we need to integrate both the spatial and temporal dynamics of signaling in cells.  \textit{Partial differential equation} (PDE) models allow us to consider the spatial aspects of biological signaling networks. Reaction-diffusion models, in particular, are ubiquitous in biology, particularly in systems which give rise to spatial patterning \cite{maini1997spatial,meinhardt1999orientation,kholodenko2006cell,jilkine2007mathematical}. 

PDE models often exhibit a rich behavior as parameters are varied, making analysis of these systems challenging. Despite their seeming simplicity, PDE models prove difficult to analyze due to their rich behavior across parameter space, especially in comparison to the phase-space of their reaction-only ODE counterparts. This daunting complexity often leads to these models being studied primarily through simulations, which provide instant but transient gratification: while we gain insight into the behavior of these systems, the variety of these behaviors across parameter space nearly assures that we cannot grasp their full capabilities through simulation alone.

Several computational tools have arisen over the past years to simulate complex biological models \cite{blatt1997pdecon,dhooge2003matcont,salinger2005bifurcation,funahashi2008celldesigner}. In the category of analytical methods, nonlinear analysis provides a comprehensive global picture of the system, but can be very difficult to implement even for simple models. A recently proposed method, local perturbation analysis (LPA), gives information about stimulus-driven cellular responses (we point the interested reader to \cite{holmes2015local} for more details on this method). However, linear (Turing) stability analysis remains the gold standard for understanding spontaneous pattern formation in reaction-diffusion systems. Despite its well-accepted usefulness in exploring such systems, there are few user-friendly tools available to the larger biological community to utilize these analytical methods. In the following, we provide an \textit{analytic} protocol for the study of reaction-diffusion systems. We demonstrate this protocol using two simple, classic biological examples of pattern formation and cell polarization. We also note that the method outlined is easily generalizable to any signaling network.

\begin{figure*}[t]
\centering
\includegraphics[width=0.9\textwidth]{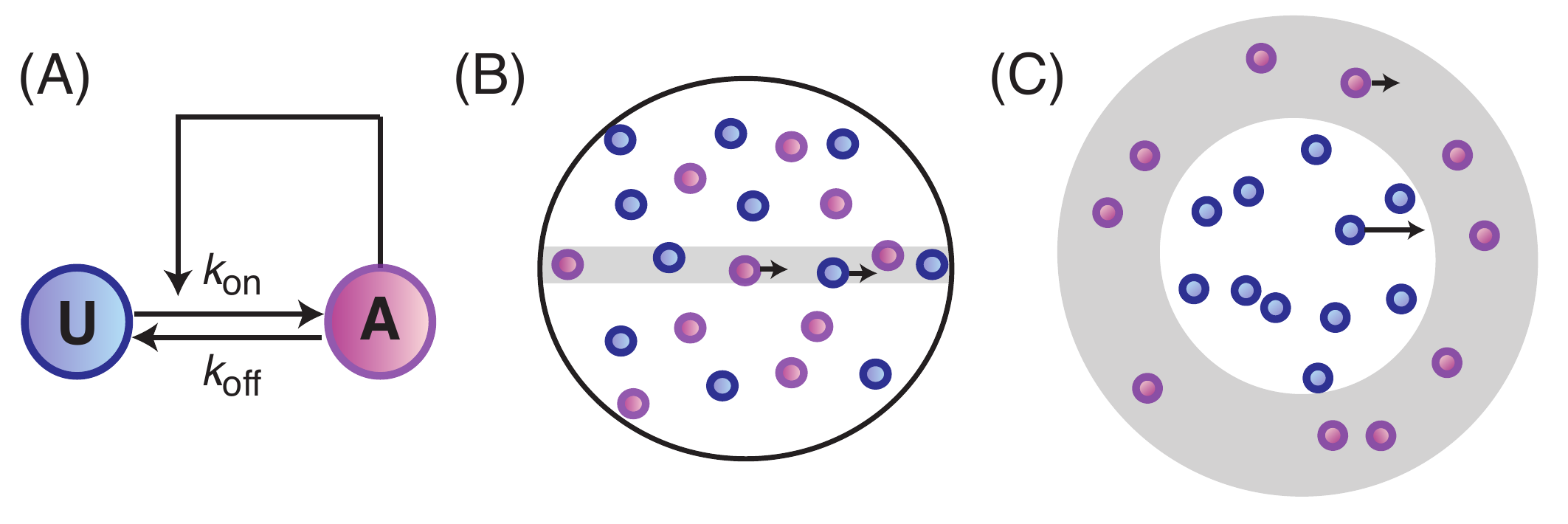}
\caption{Schematic of a simplified signaling model. \textbf{(A)} Purely nonspatial model corresponds to a well-mixed system. The model consists of an interconversion between inactive (U) and active (A) forms of a single protein. The active form effects a positive feedback on its own production; many signaling networks take the form of a cooperative Hill function. An extended one-dimensional spatial model can be representative of \textbf{(B)} two cytoplasmic species or \textbf{(C)} one cytoplasmic and one membrane-bound species, depending on the relative diffusion coefficients of the forms.}
\label{fig:schematic}
\end{figure*}

\section{General model framework}
\subsection{Reaction-diffusion equations}

We consider chemical reactions occurring along a one-dimensional segment ($0 \leq x \leq L$), which serves as an approximation of a transection of the cell. We focus primarily on examples corresponding to two chemical reactants (Figure \ref{fig:schematic}A), but note that the method described (as well as our provided computational tool) is easily extensible to systems with an arbitrary number of species. Depending on the relative values of the diffusion coefficient, this extension can serve to model several cytoplasmic species (Figure \ref{fig:schematic}B) or a mixture of cytoplasmic and membrane-bound species (Figure \ref{fig:schematic}C). Here, we provide a general example with two species U and A. We denote the concentrations of these species as $u$ and $a$ respectively, both in units of molecules/length. The partial differential equations are:
\begin{align}
\frac{\partial a}{\partial t} &= D_a\frac{\partial^2 a}{\partial x^2} + f(a,u)\nonumber\\
\frac{\partial u}{\partial t} &= D_u\frac{\partial^2 u}{\partial x^2} + g(a,u).
\label{generalpdes}
\end{align}
The choice of functions $f(a,u)$ and $g(a,u$) is arbitrary. We will first outline the general analysis method and then apply it to two specific systems to demonstrate its usefulness. 

\subsection{Stability analysis for the well-mixed system}

The first step is to find \textit{steady state} solutions $(a^*,u^*)$ of the purely kinetic system (in the case of homogenous spatial conditions), we solve for $(f(a,u), g(a,u)) = (0,0)$. Additionally, to enforce mass conservation, we impose the constraint $p = u_0 + a_0 = u(t) + a(t)$, where $p$ is a constant denoting the total amount of protein. Depending on the form of $f(a,u)$ and $g(a,u)$, multiple steady states can exist. 

These \textit{homogenous equilibria} also serve as steady states for the spatially extended system. We are interested in their stability throughout the parameter space. Next, for the spatially homogenous system, we look at the eigenvalues of the \textit{Jacobian matrix} to analyze the linear stability of these states under small perturbations from equilibrium  \cite{strogatz2014nonlinear}. The Jacobian matrix for the purely kinetic system is 
\begin{equation}
 J = \left( \begin{array}{cc} 
f(a,u)_a &  f(a,u)_u\\ 
g(a,u)_a & g(a,u)_u\end{array}
\right). 
\end{equation}
Here, $f_a$ and $f_u$ denote the partial derivatives of $f(a,u)$ with respect to $a$ and $u$, respectively. The linearization of the reaction terms around an equilibrium point $(a^*,u^*)$ is simply $J|_{(a^*,u^*)}[da \ \  du]^T$, where $J|_{(a^*,u^*)}$ represents the Jacobian matrix evaluated at $(a^*,u^*)$.  An equilibrium point is said to be stable if the eigenvalues of the $J|_{(a^*,u^*)}$ all have real parts less than or equal to zero and unstable if any one eigenvalue has a real part greater than zero. For example, in bistable systems such as MAP kinase, there are two stable steady states and one unstable steady state.

\subsection{Stability analysis for the spatial model}
What happens when we consider a spatially heterogeneous system? In order to obtain the spatially heterogenous solutions, we linearize Equations \eqref{generalpdes}, we arrive at
\begin{equation}
\frac{\partial}{\partial t}\begin{bmatrix}\partial a \\ \partial u \end{bmatrix} = \begin{bmatrix}D_A & 0 \\ 0 & D_U \end{bmatrix}\frac{\partial^2}{\partial x^2}\begin{bmatrix}\partial a \\ \partial u \end{bmatrix}  + J \begin{bmatrix}\partial a \\ \partial u \end{bmatrix}.
\label{eq:linspat}
\end{equation}
Here, $J$ is the Jacobian for the (spatially homogenous) reaction equations. In order to analyze the stability of this system with respect to perturbations, we must first note that now these perturbations depend both on time and space. A convenient form for such perturbations is $[\partial a \ \ \partial u] = [\partial a^* \ \ \partial u^*] e^{-\lambda t} e^{ikx}$, where the term $e^{ikx}$ is a common way of representing a spatial wave, with $k$ as the wavenumber (we refer the interested reader to \cite{strogatz2014nonlinear}). Substituting this into the linearized Equation \eqref{eq:linspat} leads to the Jacobian of the spatially extended system 
\begin{equation}
 J^* = \left( \begin{array}{cc} 
f(a,u)_a - D_Ak^2 &  f(a,u)_u\\ 
g(a,u)_a & g(a,u)_u-D_Uk^2\end{array}
\right). \end{equation}

The eigenvalues of this matrix allow us to study the stability properties of the system: nonnegative eigenvalues correspond to a loss of stability of the system. Since the eigenvalue expressions contain the unknown variable $k$, we must consider a \textit{one parameter family of solutions}, one for each wavenumber. Note that the eigenvalues for the well-mixed system are achieved when $k = 0$. Primarily, we are interested in the emergence of heterogenous patterns that occur via perturbations within a finite range of critical wavenumbers $0 \leq k_c \leq k_{\text{max}}$. 

The value of these critical wavenumbers becomes important when we are faced with a finite domain size. Since cells are of finite size, we want to investigate how the length of the domain affects the system response. This is because the spatial pattern that emerges from an instability corresponding to a wavenumber $k$ has wavelength $\omega = \frac{2\pi}{k}$; accordingly, for finite systems, only values of $k$ above a certain threshold will generate any meaningful spatial patterns.

We emphasize that the steps discussed in this article are highly generalizable, and we aim to highlight how such comprehensive treatments of mathematical models in biology are powerful and insightful. 

\section{Examples}
In this section, we demonstrate the above analytical protocol using two examples, each representing a distinct origin of spatial patterning. First, we discuss Turing's ``diffusion-driven instability'' using morphogenesis as a classical example. Second, we consider a situation in which spatial patterns arise even though the criteria for Turing instabilities are not met, via the ``pinning'' of a traveling wave. To illustrate this phenomena, we turn to a simple model of eukaryotic cell polarization.

\subsection{Classic Turing spots}

Morphogenesis, the process by which an embryo gains its structure during development, is one of the most studied problems 	in biology. In particular, Turing's model of pattern formation, originally applied to the development of animal coat patterns \cite{turing1952chemical}, proved useful for describing this phenomenon. Murray \cite{murray2001mathematical} used and analyzed the following system of non-linear reactions, which has since become the classical example of the Turing instability. In this system, $f(a,u)$ and $g(a,u)$ are described as follows:
\begin{eqnarray}
f(a,u) &= b - a - \frac{\rho a u}{1+ a + Ka^2}\\
\nonumber
g(a,u) &= \alpha (c-u) - \frac{\rho a u}{1+ a + Ka^2},
\label{eq:Turing}
\end{eqnarray}
where $b$, $\rho$, $K$, and $c$ are reaction rate parameters. 

\subsubsection{Well-mixed model}

We first begin with the (relatively) easier task of analyzing the well-mixed model before considering the full reaction-diffusion system. This system has two homogenous steady states, which are obtained by solving $f(a,u)=0$ and $g(a,u)=0$. We denote them as $(a_1^*,u_1^*)$ and $(a_2^*,u_2^*)$.  We do not explicitly write out the expressions for the sake of brevity.

Next, we construct the Jacobian of this system to obtain
\begin{equation}
 J = \left( \begin{array}{cc} 
\frac{Ka^2\rho u}{(1+a+Ka^2)^2}-\frac{\rho u}{1+a+Ka^2}-1 & \frac{-a\rho}{1+a+Ka^2}\\ 
\frac{Ka^2\rho u}{(1+a+Ka^2)^2} - \frac{\rho u}{1+a+Ka^2} & -\alpha -\frac{a \rho}{1+a+Ka^2}  \end{array}
\right). 
\end{equation}
The eigenvalues of this matrix have negative real parts indicating that the purely kinetic system is stable. 
\subsubsection{Spatial patterns}

\begin{figure}[!!h]
\centering
\includegraphics[width=0.45\textwidth]{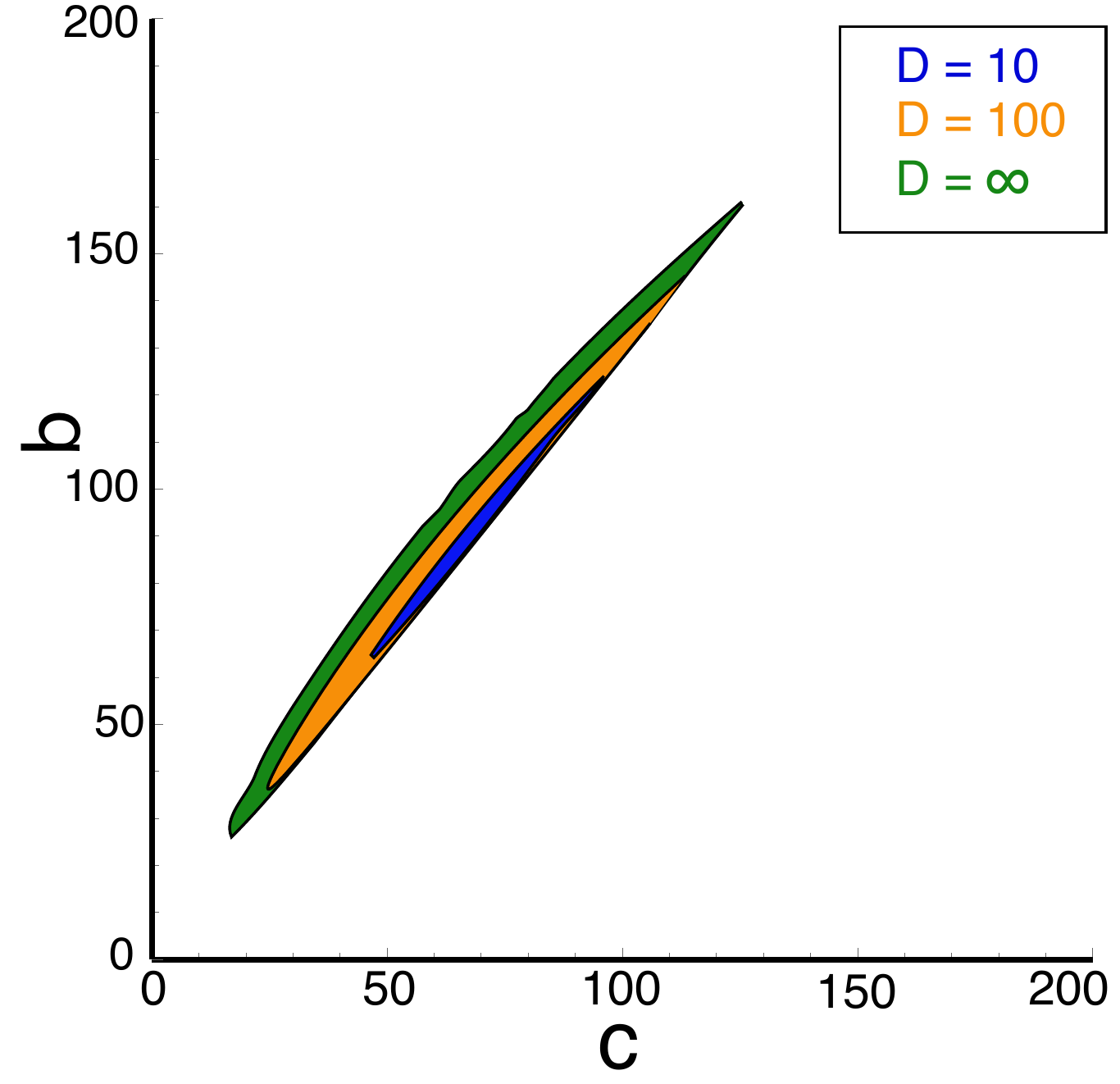}
\caption{Region of parameter space in the Thomas system (Eq. \ref{eq:Turing}) that allows for Turing patterns for the ratio of diffusion constants $D = D_a/D_u$: $D=10$ (blue), $D=100$ (orange) and $D=\infty$ (green). The following parameters are fixed as follows: $\alpha = 1.5, p = 13, K = 0.125$. As the $D$ increases, the region over which Turing patterns can be observed increases.}
\label{thomas}
\end{figure}

The criteria for an equilibrium point to undergo a diffusion-driven instability are as follows. 
\begin{itemize}
\item First, the equilibrium point must be stable for the purely kinetic system and the loss of stability in the steady state must be purely spatially dependent. 

\item Second, the Jacobian of the spatially extended system must either have a trace less than 0 or a determinant greater than 0. That is, the system \textit{cannot} undergo a classical Turing instability if both:
\begin{eqnarray}
\text{tr} (J^*) =& f_a + g_u < 0,\\
\label{eq:conditions_turing}
\nonumber
\text{det} (J^*) =& f_ag_u - f_ug_a > 0.
\end{eqnarray}

For the above system, the conditions for the Turing mechanism are met, and spatial patterning through a diffusion-driven instability can occur. Note that there are six parameters in the system, including $D = D_a/D_u$, $b$, $c$, $K$, $\alpha$, and $\rho$. In Figure \ref{thomas}, we fix $K$, $\alpha$, and $\rho$ and study the effect of $D$ on the phase space admitting Turing patterns over the ranges of parameters $b$ and $c$. When $D=10$, the range of $b$ and $c$ over which Turing patterns are observed is small. However, as $D$ increases, the range of $b$ and $c$ increases. Biologically, this means when A diffuses much faster than U, the range of parameters $b$ and $c$ over which Turing patterns can be observed increases.  For more details on linear instabilities via the Turing mechanism, we refer the interested reader to \cite{murray2001mathematical}.

\end{itemize}

\subsubsection{Key takeaway points}

Classical Turing systems consist of two components, usually an activator and an inhibitor  . Since Turing's original paper on pattern formation via a reaction-diffusion mechanism, several models have been proposed that are applicable to a wide variety of biochemical problems  . 

Above, we have described and analyzed Thomas' model for substrate inhibition based on experiments in which an immobilized enzyme on a membrane react with diffusing substrate and co-substrate molecules \cite{thomas1976}. This model, and several others proposed after it \cite{seelig1976,catalano1981} help us understand biologically plausible mechanisms of signaling. Some of the key points to note from studies of Turing patterns are noted below.

\begin{enumerate}

\item{\textbf{Specific kinetics are required for Turing patterns.} Two component mechanisms, like the one described in Eq. \ref{eq:Turing}, can generate spatially heterogenous patterns. Whether or not systems are capable of generating spatial patterns through a Turing mechanism depends on the reaction kinetics. In particular, we have outlined the conditions required in Eq. \ref{eq:conditions_turing}. In general, these systems crucially have an activation-inhibition form (Figure \ref{fig:schematic}A). Further details on such mechanisms can be found in \cite{murray2001mathematical}.}

\item{\textbf{Diffusion can lead to instabilities under certain conditions.} Given that the kinetics fulfill the conditions outlined above, Turing instabilities can arise only if the ratio of diffusion constants, $D=D_a/D_u \neq 1$. These spatial heterogeneities arise especially when the diffusion constant of the activator in the system is smaller than that or the inhibitor (called LALI, or "local activation, lateral inhibition"	 mechanisms). The concept of diffusion-driven instability was a groundbreaking concept when it was first proposed because diffusion was long-considered to be a stabilizing presence. The patterns that arise from these instabilities are called Turing patterns, and are thought to commonly arise in several biological systems.}

\end{enumerate}

\subsection{Cell polarization}
Polarization is fundamental to eukaryotic motility, morphogenesis, and division  . Cell polarization is the process of reorganization of the cytoskeleton into distinct front and back regions in response to a stimulus  . Recently, Mori et al. \cite{mori2008wave} presented the `wave pinning polarization' or WPP model, a minimal model of cell polarization, that demonstrated both bistability and spatial patterning. In this system, the kinetics of the two species are given by:
\begin{eqnarray}
f(a,u) = u\left(k_0 + \frac{\gamma a^2}{K^2 + a^2} - k_{off} a\right)\\
\nonumber
g(a,u)  = -f(a,u)=-u\left(k_0 + \frac{\gamma a^2}{K^2 + a^2} - k_{off} a\right). 
\end{eqnarray}

After their initial proposal of the WPP model \cite{mori2008wave}, the authors considered several important extensions \cite{mori2011asymptotic,jilkine2011comparison,walther2012deterministic}. However, a full characterization of the equilibrium behavior of the system as a function of the basal rate of activation $k_0$ and the average amount of total protein $p$ has not yet been adequately discussed. In this section, we outline such a characterization. The mathematical basis of such analyses can be found in the fundamental book by Strogatz \cite{strogatz2001nonlinear}. Computations were done in Mathematica and MATLAB; all associated scripts are available as a user-friendly GUI at www.ocf.berkeley.edu/$\sim$jnirody$^{\dag}$.

\subsubsection{Well-mixed model}

As before, we begin by analyzing the reaction-only ODE model before considering the full reaction-diffusion system. The WPP model has three steady states, which we denote $(a_1^*,u_1^*)$,  $(a_2^*,u_2^*)$, and $(a_3^*,u_3^*)$.  To assess the stability of each of these steady states, we look at the \textit{Jacobian matrix} of the system:
\begin{equation}
 J  =  \left( \begin{array}{cc} 
\frac{2auK^2\gamma}{(a^2+K^2)^2} - k_{\text{off}} & k_0 + \frac{\gamma a^2}{K^2+a^2}\\ 
- \frac{2auK^2\gamma}{(a^2+K^2)^2} + k_{\text{off}} & -k_0 - \frac{\gamma a^2}{K^2+a^2} \end{array}
\right). 
\end{equation}

\begin{figure}[t]
\centering
\includegraphics[width=\textwidth]{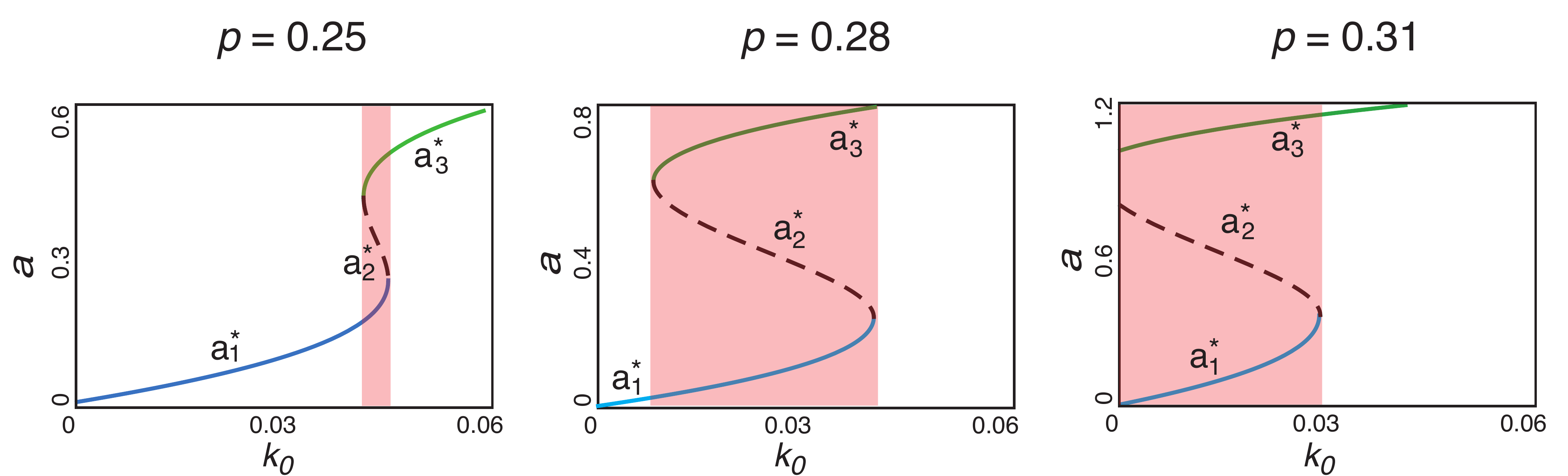}
\caption{Equilibrium curves for each of the three steady states in the WPP model. For all of the following plots, we choose: $\gamma=1$, $k_{\text{off}}=1$, $K=1$ and vary the average amount of total protein $p$. The two stable steady states $a_1^*$ and $a_3^*$ are shown as solid blue and green lines, respectively; the unstable steady state $a_2^*$ is shown as a dashed line. For a range of $k_0$, all three steady states exist and are real-valued; this region is shaded in red; this range increases with $p$, eventually resulting in an \textit{irreversible} system response when it reaches $k_0=0$. }
\label{fig:eqcurves}
\end{figure}

Example equilibrium curves for $\gamma = 1, k_{\text{off}} = 1, K =1$ are shown for various values of average total protein $p$ in  Figure \ref{fig:eqcurves}. For the WPP model, $a_1^*$ and  $a_3^*$ are stable (shown as blue and green solid lines, respectively), while $a_2^*$ is unstable (shown as a black dashed line). For certain choices of parameters, both stable steady states exist; this is called the \textit{bistable regime}. Within this region, either of the two steady states may be reached for the same set of kinetic parameters, depending on the initial conditions of the system (i.e., the `starting' concentrations). 

Bistability is a common feature in biochemical reaction networks, particularly those containing positive feedback loops \cite{xiong2003positive}. Through bistability, positive feedback loops may allow for a sustained cellular response to a transient external stimulus \cite{xiong2003positive}, a central feature in cell polarization. To illustrate this, consider the basal rate parameter $k_0$ as a function of an external stimulus $S$ (i.e., $k_0 = k_0^*S$). Now, we can look to the plots in Figure \ref{fig:eqcurves} as dose-response curves --- the cell responds to external stimulus $S$ by producing the activated protein A.

As $S$ (and consequently $k_0$) is slowly increased, the concentration of A follows along the curve corresponding to $a_1^*$ (blue) until it crosses the bistable region, after which the equilibrium value is the larger $a^*_3$ (green). If the stimulus is removed, and the level of $S$ decreases, the higher-valued equilibrium is maintained within the bistable region; this behavior, where the dose-response relationship is in the form of a loop rather than a curve, is called \textit{hysteresis}. If the bistable regime is large enough, in particular if it extends to $k=0$, an essentially irreversible response to transient stimuli may be elicited (as is seen for $p = 0.31$ in Figure \ref{fig:eqcurves}).

\begin{figure}[t]
\centering
\includegraphics[width=0.4\textwidth]{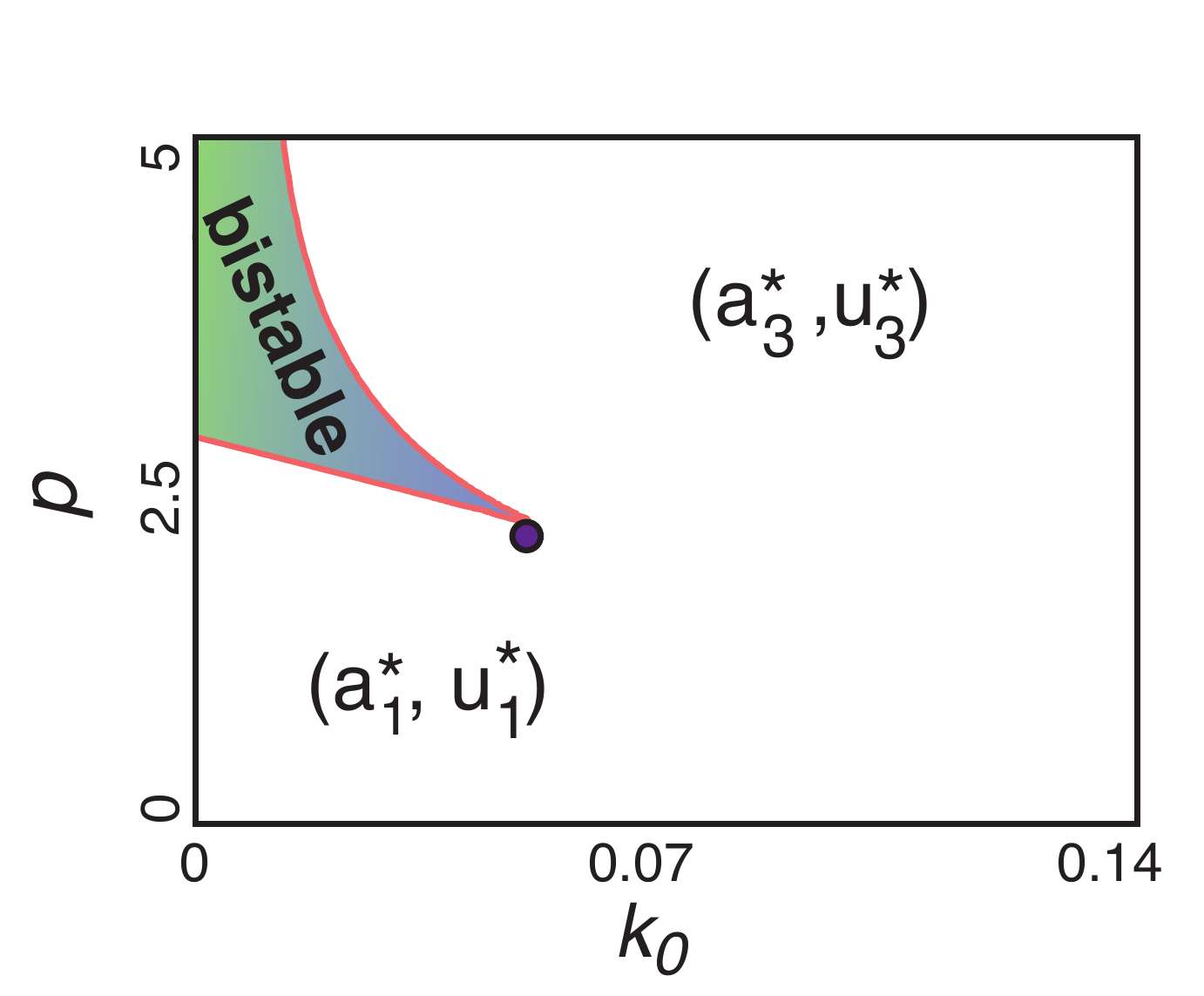}
\caption{The extent of the bistable region in $(k_0,p)$ space for the well-mixed model. In unshaded regions, only a single stable steady state (either $(a_3^*,u_3^*)$ or $(a_1^*,u_1^*)$) exists. Within the shaded region, both steady states exist and are stable. The system may tend to either state, dependent on initial conditions. The parameter set used in the simulations by Mori \emph{et al.} \cite{mori2008wave} is shown as a purple point.}
\label{fig:homdiag}
\end{figure}

Figure \ref{fig:homdiag} shows the bistability region for the WPP model as the total average protein concentration $p$ and the basal activation rate $k_0$ are varied. From the sloping shape of the bistable region, we can see that for sufficiently high values of total protein (for the set of parameters in Figure \ref{fig:homdiag}, $p \geq 3$), an `irreversible' response may be generated: both steady states are stable even when the stimulus is removed, at $k_0 = 0$. The simulation parameters chosen by Mori \emph{et al.} are shown as a purple dot.

\subsubsection{1-D spatial model}

Having characterized the bistable regime of the well-mixed model, we now turn our attention to the full reaction-diffusion system. The homogenous equilibria also serve as steady states for the spatially extended system and as  before, we are interested in their stability throughout the parameter space. 

Recall that for the spatially homogenous system, we turned to the Jacobian to analyze the linear stability of these states under small perturbations from equilibrium. The eigenvalues of $J$ are
\begin{equation}
\sigma^{\pm} = \frac{1}{2}\left[\left(f_a - f_u - (D_A + D_U)k^2\right) \pm 
 \sqrt{\left(f_a - f_u - (D_A + D_U)k^2\right)^2 - 4 \left(D_AD_Uk^4 + (D_Af_u - D_u f_a)k^2\right)}\right].
\end{equation}
As the expressions for the eigenvalues now contain the additional unknown $k$, we are now interested in a \textit{family of solutions}, one for each wavenumber. The eigenvalues for the nonspatial system are achieved when $k = 0$, and are given by $\sigma^- = f_a - f_u$ and $\sigma^+ = 0$; therefore any spatially homogenous perturbation will relax back to the spatially uniform steady state. We are instead interested in the emergence of heterogenous patterns that occur via perturbations within a finite range of critical wavenumbers $0 \leq k_c \leq k_{\text{max}}$. 
\begin{figure}[h]
\centering
\includegraphics[width=0.45\textwidth]{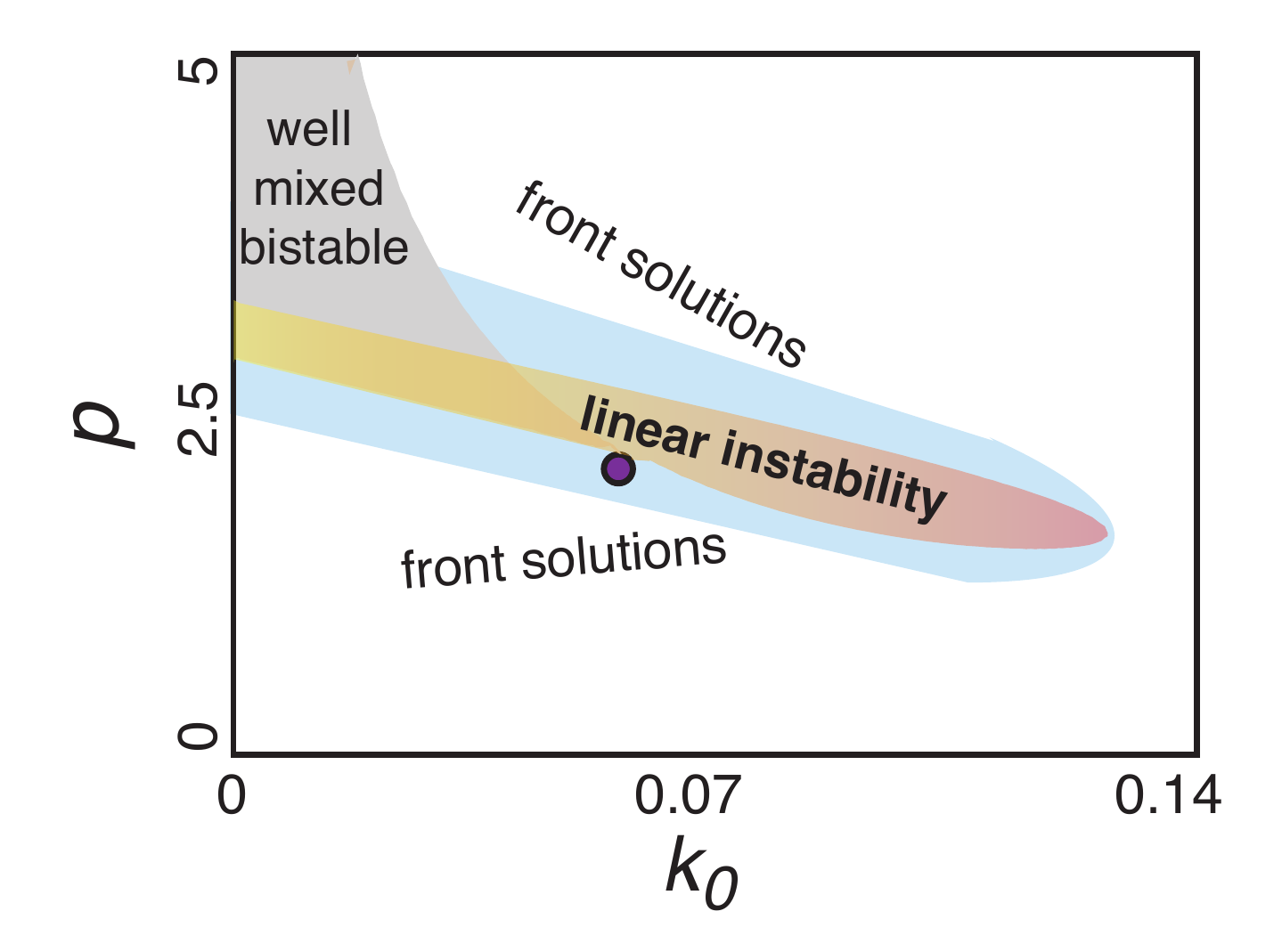}
\caption{Parameter space topology for the full PDE model when $D_U = 10\mu$m$^2$s$^{-1}$  and $D_A = 0.1\mu$m$^2$s$^{-1}$. The region of linear instability is shown shaded in orange for wavenumber $k = 0.2$ $\mu m^{-1}$. This  corresponds to a perturbation of length $L=\frac{2\pi}{k} \approx 30 \mu$m. Smaller values of $k$ result in an expansion of the linear instability region; larger values of $k$ result in the region shrinking. An additional domain is shown shaded in blue, in which front-like solutions are supported when given a sufficiently strong (or spatially graded) perturbation. The parameter choice made by Mori \emph{et al.} (purple point) lies in this region.}
\label{fig:hetdiag}
\end{figure}

\begin{figure*}[t]
\centering
\includegraphics[width=0.85\textwidth]{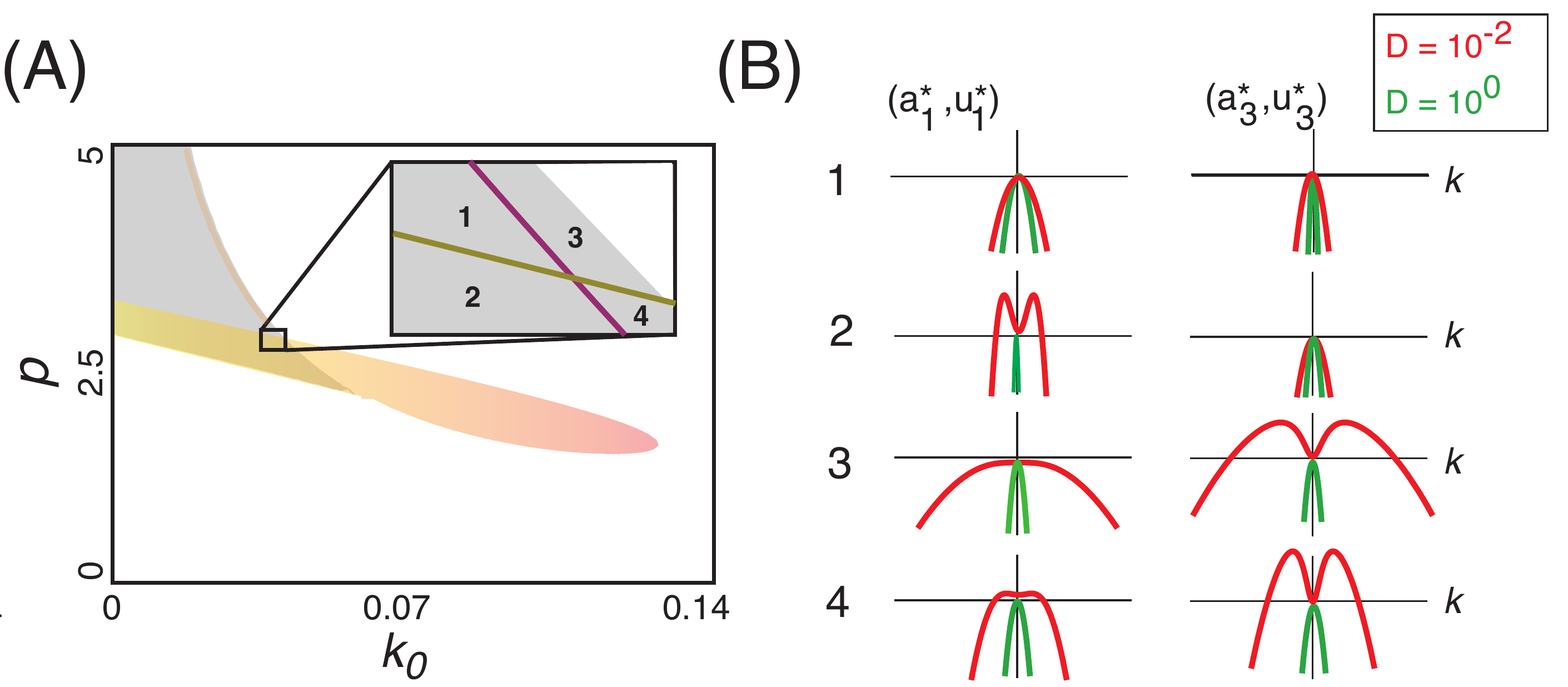}
\caption{Examination of the rise of Turing patterns by loss of stability given critical wave numbers $k$. \textbf{(A)} A small segment of the parameter space is highlighted for the full spatial WPP model, showing regions where neither (region 1), one (regions 2 and 3), or both (region 4) of the equilibrium points become unstable. \textbf{(B)} An illustration of the loss of the linearly instability regime as $D$ approaches 1. Plots show the magnitude of the real part of the rightmost eigenvalue for both equilibria within each of the regions highlighted in \textbf{A}. Plots are shown for $k \in [-1,1]$. When $D = 10^{-2}$, corresponding to the localization of the active form to the membrane, a finite range of critical wavenumbers is observed; this range disappears when $D = 1$.}
\label{fig:wavenumbers}
\end{figure*}

The region of `linear instability' is highlighted in orange in Figure \ref{fig:hetdiag} for $k=0.2$ $\mu m^{-1}$. This corresponds to a pattern of wavelength $\omega = \frac{2\pi}{k} \approx 30\mu$m. This length scale is intermediate among the motile eukaryotic cells that use Rho GTPases to generate polarity. In this region, one or both of the steady states, $(a_1^*,u_1^*)$ and $(a_3^*,u_3^*)$, lose stability with respect to a inhomogeneous perturbation as shown above (see Figure \ref{fig:wavenumbers}).

We note that the region computed is for diffusion coefficients $D_U~=~10~\mu$m$^2s^{-1}$ and $D_A = 0.1$ $\mu$m$^2s^{-1}$. A similar parameter topology (with slightly larger regions of instability) was found in a reduced one-species model where infinite cytoplasmic diffusion was assumed \cite{trong2014parameter}. The disparity in diffusion coefficients assumed in Figure \ref{fig:hetdiag} presupposes the compartmentalization of the two species: a protein diffuses far more slowly on the membrane than in the cytosol (here, we assume the ratio of diffusion coefficients to be $\approx 0.01$ \cite{postma2004chemotaxis}). In the following sections, we will assess the importance of this presumption.
In addition to the region of linear instability, the parameter space for the full model can admit a surrounding region where front-like solutions are observed. In this region, a stalled wave can appear when the system is subjected to a directed stimulus (e.g., a gradient) or if the domain exhibits some intrinsic polarity at $t=0$. The parameters chosen by Mori \emph{et al.} fall within this region; their simulations demonstrate that this `intrinsic polarity' may arise via sufficiently noisy initial conditions \cite{mori2008wave}.

The boundaries of this regime are calculated by solving the Maxwell condition and finding the ranges of $u$ that admit a stalled wave \cite{mori2008wave}:
\begin{equation}
I(b) = \int_{a_-}^{a_+} f(a,u)da = 0.
\end{equation}  
We then use the mass conservation condition to compute this region in $(k_0,p)$ space. While this is not analytically feasible for the WPP model, numerically solving the integral provides an accurate characterization of this region. The method outlined here is not specific to the WPP model, or even to models of cell polarity. To make such a phase-space analysis more accessible to the biological community, we provide an easy-to use GUI to allow readers to analyze the stability properties of other systems of interest.

\subsubsection{Key take-away points}

The analysis above provides us significant insight into the behavior of the WPP model throughout the parameter space (Figure \ref{fig:hetdiag}). Using this, we can point out several notable properties of the model as they pertain to cell polarization.

We note that the below properties are predicted using analyses performed on a 1D spatial model. Extension of the model to three, or even two, dimensions may (and likely do) result in different behaviors \cite{edelstein2013simple}. However, an analytic treatment of higher-dimensional spatial models is rarely possible, and thus the comprehensive analysis of a reduced model in one spatial dimension lays the foundation for simulation studies in higher dimensions.

\begin{enumerate}
\item{\textbf{Importance of compartmentalization:} The Rho GTPase family is large and varied, and is present in eukaryotes spanning from \textit{C. elegans} to humans. However, one common feature of these proteins is \textit{compartmentalization}: the active form is bound to the membrane, while the inactive form diffuses in the cytoplasm \cite{raftopoulou2004cell}. This feature has been shown to be important for cell polarization \cite{mori2008wave,jilkine2011comparison,mori2011asymptotic}. We illustrate the necessity of membrane localization by considering how the phase space of the WPP model in Figure \ref{fig:hetdiag} changes if both species are contained in the cytoplasm.

Qualitatively, the dependence of polarization on compartmentalization is relatively intuitive. As A and U constitute GTP- and GDP-bound versions of a single protein, their cytoplasmic diffusion rates are likely very similar. Given this, one would not expect the formation of any sort of regular pattern with no initial spatial structure. 

We show this quantitatively by defining the ratio of the diffusion rates $D = D_A/D_U$, and considering the effect of this quantity on the regimes allowing spatially heterogenous solutions. For one membrane-bound and one cytoplasmic species, we take this ratio to be $\approx 0.01$ \cite{postma2004chemotaxis}. 

When $D = 1$, no spatially heterogenous patterns can be generated, and this region disappears altogether. However, as $D$ decreases, spatial patterns are supported for a finite range of wavenumbers (Figure \ref{fig:wavenumbers})  \cite{mori2011asymptotic}. As the rate of diffusion of U increases, the maxima of $\sigma(k)$ move towards $k=0$, eventually displacing the uniform steady state in the limit $D_U \rightarrow \infty$ \cite{trong2014parameter}. This trend is not symmetric --- as $D$ increases from 1, there is no consequent extension of the multistable regime. This is similar to the formation of Turing patterns in local excitation, global inhibition models; for an overview of this brand of models with respect to cell polarization, we refer the reader to several excellent reviews \cite{mori2008wave,jilkine2011comparison,edelstein2013simple}. }
\item{\textbf{Benefit of being big:} 
In addition to the importance of different diffusion rates between active and inactive forms, we can use the fact that the range of critical wave numbers is bounded above to consider the existence of a corresponding \textit{lower bound} on the length of the cell $L$.  

The value of the maximum critical wavenumber $k_{\text{max}}$ for a feasible value of $D$ is quite low (Figure \ref{fig:wavenumbers}). This suggests that smaller cells are less sensitive to polarization, while larger cells are able to respond more robustly. Interestingly, this result has been observed experimentally: cells were found to become significantly more sensitized as they were flattened in a confined channel \cite{meyers2006potential,holmes2012modelling}. }
\item{\textbf{Spontaneous polarization:} Recall that $k_0$ can be written as a function of the concentration of some stimulus $S$: $k_0(S) = k_0^*S$. This formulation allows us to characterize parameter values which allow for \textit{spontaneous polarization} in the absence of a directed external stimulus.

Certain, but not all, cells are able to spontaneously self-polarize. In the WPP model, we can by considering the regimes crossed in Figure \ref{fig:hetdiag} by the line $k_0 = 0$. We see that polarization in the absence of a stimulus is possible if the value of $p$ is sufficiently high. This is consistent with experimental observations that the constitutive expression of Rho GTPases result in extension of randomly oriented lamellopodia and membrane ruffling \cite{rei1996phosphatidylinositol}.  } 
\item{\textbf{Polarization strategies: } The parameter space topology for the WPP model contains two distinct regions that allow for non-homogenous equilibrium solutions. Because of the choice of parameters in Mori \emph{et al.}\cite{mori2008wave}, the system behavior in only one of these regions, corresponding to stalled-wave solutions (shaded in blue in Figure \ref{fig:hetdiag}) was explored. The fact that these solutions were not initially found by a purely simulation-based study further emphasizes the need for comprehensive analytic treatment of biological models.

In general, Turing patterns form more easily (i.e., in response to far smaller perturbations) than patterns formed by a wave-pinning mechanism. However, they occur on a far slower timescale \cite{mori2008wave,jilkine2011comparison,edelstein2013simple}. The existence of a Turing-like instability regime in addition to a region which admits stalled-wave solutions presents cells with multiple strategies for polarization.}
\end{enumerate}

\section{Conclusions}
Mathematical modeling has served to complement experiment to further our understanding of biological systems. Recent advances in knowledge of these systems have allowed us to construct more and more complex models, but our computational abilities have not caught up. In particular, our knowledge of the cell has made us increasingly aware of the structured nature of the cellular environment. 

Despite this knowledge, biological models often neglect this structure and consider biological networks in a homogenous spatial environment. These purely kinetic models are computationally more tractable, but can fail to capture the biological reality. However, incorporating spatial structure into these mathematical models can make them difficult to analyze using current tools.

For systems involving spatial patterning, such models are often of the reaction-diffusion type. As the behavior of these systems can vary widely with parameter choice, simulation studies are often unsatisfying in that they only provide a small peek into the full capabilities of these models. Above, we have outlined the steps for a comprehensive analytic treatment of spatial structure arising from reaction-diffusion models. 

Additionally, we  described examples from two distinct patterning mechanisms: diffusion-driven instability (also known as the Turing mechanism) and wave-pinning. We have highlighted the usefulness of such a protocol, in that it illuminates several features of models which are difficult to observe via simulation alone.

\footnotesize{
\bibliography{IB_ref} 
\bibliographystyle{plain} 
}

\end{document}